\documentclass[12pt]{article}
\usepackage{latexsym,epsfig}
\usepackage{graphicx}
\usepackage{pifont}
\usepackage{verbatim}
\def\ed{\end{document}}

\def\sk{\smallskip}
\def\md{\medskip}

\def\beq{\begin{eqnarray}}
\def\eq{\end{eqnarray}}
\def\beqn{\begin{eqnarray*}}
\def\eqn{\end{eqnarray*}}
\def\nl{\noindent}

\begin{document}
\begin{center}
\vglue - 1 cm
{\bf \Large Top quark forward-backward asymmetry }
\md

{\bf \Large from the  $3-3-1$  model}
\md

{\bf }
\md

{E. Ramirez Barreto}

{Centro de Ci\^encias Naturais e Humanas, UFABC}

{Santo Andr\'e, SP, Brazil}
\vskip .3 cm

{Y. A. Coutinho }

{Instituto de F\'isica, UFRJ
}

{Rio de Janeiro, RJ, Brazil}
\vskip .3 cm

{J. S\'a Borges}

{Instituto de F\'isica, UERJ
}

{Rio de Janeiro, RJ, Brazil}
\end{center}
\sk

\begin {abstract}

The forward-backward asymmetry $A_{FB}$ in top quark pair production, measured at the Tevatron,  is probably related to the contribution of new particles.
The Tevatron  result is more than a $2\sigma$ deviation from the standard model prediction and motivates the application of alternative models
introducing new states. 
 However, as the standard model predictions for the total cross section $\sigma_{tt}$  and invariant mass distribution $M_{tt}$ for this process are 
in good agreement with experiments, any alternative model must reproduce these predictions. These models can be placed into two categories:
One introduces the s-channel exchange of new vector bosons with chiral couplings to the light quarks and to the top quark and  
another relies on the t-channel exchange of particles with large  flavor-violating couplings in the quark sector.
In this work we employ a model which introduces both s- and t-channel  nonstandard contributions for  the 
top quark pair production in proton antiproton collisions. We use the minimal version  of the $SU(3)_C \otimes SU(3)_L 
\otimes U (1)_X$ model (3-3-1 model) that 
predicts the existence of a new neutral gauge boson,  called $Z^\prime$. This gauge boson has both flavor-changing couplings to up and top quarks and chiral coupling  to the light quarks and to the top quark. This very peculiar model coupling can correct
the $A_{FB}$ for top quark pair production for two ranges of $Z^\prime$ mass while leading to cross section and invariant mass distribution quite similar to the standard model ones. This result reinforces the role of the 3-3-1 model for any  new physics effect. 
\end {abstract}

\nl PACS numbers 14.65.Ha,  14.70.Pw
\newpage
 \vglue - 2 cm
\section{Introduction}

The experimental results from  LEP, SLD (Slac Large Detector) and the Tevatron are in  accordance with all the predictions of the standard model (SM)   and, apart from the 
Higgs 
particle, all predicted particles have been  discovered. However, some indications that the SM cannot explain all of the experimental results come, 
for example, 
 from neutrino oscillations data that require different masses for the three  neutrino kinds and mixing among them. Another indication  of experimental
 difficulties faced by the SM comes from the forward-backward top quark asymmetry $A_{FB}$ measured at the Tevatron  in the top pair production process. 
On the other hand,  it is believed that the SM  is not the complete theory because it does not explain some theoretical features, for example, the 
family replication and the value of the Weinberg angle $\theta_W$. 

It would be interesting to study the top pair production process within a theoretical framework being an extension of the SM that presents an issue
 for the family replication problem and for the bound on the Weinberg angle. This is the motivation of the present work: to use the framework of the 
minimal version  of the 3-3-1 model \cite{PIV,FRA} in order to calculate the forward-backward top quark asymmetry in the  top pair production in proton-
antiproton collisions at Fermilab energy. 

The discrepancy in the $A_{FB}$ we are referring to is more than $2\sigma$ because the SM calculation, including next-to-leading order QCD corrections,
 is $A_{FB} = 5.0 \pm 1.5 \%$ \cite{STE}, while CDF obtained $ 0.158 \pm 0.075$ (stat+syst) 
in $ t \bar t$ rest frame \cite{AFBCDF} and D$0$ obtained  $8 \pm 4$ (stat)$ \pm 1 $ (syst)  $\%$ \cite{AFBD0}. 

There  are some proposed models to explain the observed discrepancies between the Tevatron result and SM calculations. A larger $A_{FB}$ than predicted by 
the SM  is obtained in models where the  {\it s-channel} exchange includes a new vector boson with chiral coupling to the light quarks \cite{MUY} or in 
models where a large  flavor-violating particle is exchanged in the  {\it t channel} \cite{SHU,CAO}. The 3-3-1 model is a possible solution for the present problem 
because it provides  the two kind of contributions, namely, a new neutral gauge boson contribution in the {\it s channel}, as well as the expected flavor-changing neutral-current (FCNC) contribution  in the {\it t channel}.
In this work we use the 3-3-1 model to calculate some distributions related to the top quark pair production in $p \bar p$ collisions.

 We present the total cross section and the invariant mass and transverse momentum distributions for three different $Z^\prime$ mass values. Our main result is a  large $A_{FB}$ while keeping other signatures of the process similar to those predicted by the SM.  The paper includes a section where the main ingredients of the model are presented followed by the results and  conclusion section.

\section{Model}

The 3-3-1 model is a gauge theory with a larger group of symmetry than the SM. It is  based on the semisimple gauge group $SU(3)_C \otimes SU(3)_L 
\otimes U (1)_X$  and,  as a consequence, it contains new gauge bosons ($W_\mu^4 ... W_\mu^8$ and $B_\mu$), fermions, and  scalars. These particles 
were until now not experimentally observed, but their existence can lead to interesting signatures. An important motivation to study the 3-3-1 model is that the predicted new particles  
are expected
 to occur at energies near the  breaking scale of the SM. Moreover this model offers an explanation of family replication following from the  
anomaly cancellation procedure 
and establishes a bound for the Weinberg angle \cite{VIP}. 

In the 3-3-1 model, the electric charge operator is defined  as  
\beq 
Q = T_3 + \beta T_8 + X I
\label {beta} \eq
\nl where $T_3$ and $ T_8$ are two of eight generators satisfying the $SU(3)$ algebra
\beq \left[ T_i\, , T_j\, \right] = i f_{i j k} T_k \quad i,j,k =1 .. 8,\eq
\nl  $I$ is the unit matrix, and $X$ denotes the $U(1)$ charge before the symmetry breaking.

The electric charge operator  determines how the fields are arranged in each representation and depends on $\beta$.  
Among the choices, $\beta = -\sqrt 3$ corresponds to the minimal version of the model, largely explored in phenomenological applications  \cite{PIV,FRA}. 
The choice
 $\beta = - 1/\sqrt 3$, which avoids exotic charged fields, leads to a version with right-handed neutrinos \cite{RHN}. 

In its minimal version (3-3-1 MIN), with $\beta=-\sqrt 3$, the  model has five additional gauge bosons beyond the SM ones. They are
 a neutral {$Z'$} and four heavy charged bileptons ${Y^{\pm\pm},V^\pm} $ with lepton number {$  L = \mp 2$}. 
\md

We define the $\gamma$, $Z$, and the new $Z^\prime$ fields as
\beq
A_{\mu}&=& s_{_W}\, W_{\mu}^{3} \, -\, \sqrt{3}\, s_{_W}\, W_{\mu}^{8}+ \sqrt{1-4\,s^2_{_W}}\, B_{\mu}, \nonumber \\
Z_{\mu}&=& c_{_W}\, W_{\mu}^{3} \,  + \sqrt{3}\, s_{_W} \, t_{_W}\, W_{\mu}^{8} \, -\,  t_{_W} \, \sqrt{1-4\,s^2_{_W}}\, B_{\mu}, \nonumber \\
Z^{\prime}_{\mu}&=& \frac{1}{c_W}\,  \sqrt{1-4\,s^2_{_W}}\, W_{\mu}^{8} \, + \sqrt 3\,  t_{_W}\,  B_{\mu},
\eq
\nl where  $c_{_W}= \cos\theta_{_W}$,  $s_{_W}= \sin \theta_{_W}$, and $t_{_W} = s_{_W}/c_{_W}$. 

Two quark families ($m=1,2$) and the third one are accommodated in $SU(3)_L$ antitriplet and triplet representation respectively: 
\begin{eqnarray}
Q_{m L} = \left( d_m \  u_m \ j_m
\right)_{L}^T \ \sim \left({\bf 3}, {\bf 3^*}, -1/3 \right),  
\quad Q_{3 L} = \left( u_3 \ d_3 \  J
\right)_{L}^T \ \sim \left({\bf 3}, {\bf 3}, 2/3 \right).
\end{eqnarray}
\nl The right-hand components of the quark fields are $SU(3)_L$ singlets:
\begin{eqnarray}
u_{\alpha R}\ \sim \left({\bf 3}, {\bf 1}, 2/3 \right)&,& \  d_{\alpha R} \ \sim \left({\bf 3}, {\bf 1}, -1/3 \right),\nonumber \\
 J_{R}\ \sim \left({\bf 3}, {\bf 1}, 5/3 \right)&,& \  j_{m R} \ \sim \left({\bf 3}, {\bf 1}, -4/3 \right),
\end{eqnarray} 
\nl $j_1$, $j_2$, and $J$ are exotic quarks with, respectively, $-4/3$, $-4/3$, and $5/3$ units of positron charge and $\alpha = 1,2,3 $, and the 
values in the 
parentheses denote quantum numbers relative to $SU(3)_C$, $SU(3)_L$, and $U(1)_X$ transformations.  
\sk

As referred to before, in the 3-3-1 model, one family must  transform  with respect to $SU(3)$  rotations  differently than the other two.  This requirement 
 manifests itself when we collect the quark currents  in a part with  universal coupling with  $Z^\prime$ similar to the SM and another part corresponding 
to the  nondiagonal  $Z^\prime$ couplings.
The transformation of these nondiagonal terms, in the mass eigenstates basis,  leads to the flavor-changing neutral Lagrangian
\begin{eqnarray}
{\cal L}_{FCNC}= \frac{g\, c_{_W}}{\sqrt{3 -12 s_{_W}^2 }}\left(\bar U_L\,\gamma^\mu \, {\cal U}^{\dagger}_L \, B \, {\cal U}_L \, U_L +  \bar D_L \, 
\gamma^\mu \, {\cal V}^{\dagger}_L \, B  {\cal V}_L \, D_L \right) Z^{\prime}_\mu.
\end{eqnarray}
\noindent  where
$$ U_L =  \left( u \,\,\,\, c \,\,\,\,  t
\right)^T_L, \quad  D_L =  \left( d \,\,\,\,  s \,\,\,\, b \right)^T_L \quad {\hbox{and}} \quad  B = {\hbox{diag}}\ \left( 1 \ 0 \ 0\right).$$
The mixing matrices  ${\cal U}$ (for {\it up}-type  quark)  and  ${\cal V}$ (for {\it down}-type quark) come from 
the Yukawa Lagrangian that gives rise to the quark masses,  and they are related to the Cabibbo-Kobayashi-Maskawa matrix as
\begin{equation}
{\cal U}^{\,u\,\dagger} {\cal V}^{\, d} = V_{CKM},
\end{equation}
In  the SM   the  {\it up}-type quark mass eigenstates are the weak  eigenstates, so \ ${\cal U}=I$. For  the 3-3-1 model, there are more states than the 
SM 
one and also $Z^\prime$ couples differently to one of the three families. In this model, the {\it up}-type  weak and the mass eigenstates 
are different inducing  $Z^\prime$ flavor changing, so ${\cal U}$ must be $ \not =  I$ \cite {LIU}. 

In the Table I, we present $Z$ and $Z^\prime$ couplings to the {\it u}- and {\it d}-type quarks. ${\cal U}$ matrix elements are free parameters; however, some limits  
for ${\cal V}$ elements have been obtained  from $Z^{\prime}$ rare decay bounds in Refs. {PRO,SHE,GOM}. In the results section, we show the values of
 the relevant matrices elements used in our calculation.

\begin{table}[ht]\label{perro}
\begin{footnotesize}
\begin{center}
\begin{tabular}{||c|c|c||}
     \hline
\hline
&    &          \\ 
&  Vector couplings & Axial-vector couplings    \\
&    &          \\ \hline
\hline
&    &      \\
$Z^{\prime} \bar c u$ & $\displaystyle{-\frac{{\cal U}^*_{12}\,{\cal U}_{11}\, \cos^2\theta_W }{\sqrt{3-12\,\sin^2\theta_W}}}$ &
$\displaystyle{\frac{{\cal U}^*_{12}\,{\cal U}_{11}\, \cos^2\theta_W }{\sqrt{3-12\,\sin^2\theta_W}}}$  \\
&      &   \\ \hline
\hline
&       &   \\
$Z^{\prime} \bar t u$  &   $\displaystyle{-\frac{{\cal U}^*_{13}\,{\cal U}_{11}\, \cos^2\theta_W }{\sqrt{3-12\,\sin^2\theta_W}}}$  &
$\displaystyle{\frac{{\cal U}^*_{13}\,{\cal U}_{11}\, \cos^2\theta_W }{\sqrt{3-12\,\sin^2\theta_W}}}$  \\
&     &     \\  \hline
\hline
&     &     \\
$Z^{\prime} \bar t c$ & $\displaystyle{-\frac{{\cal U}^*_{13}\,{\cal U}_{12}\, \cos^2\theta_W }{\sqrt{3-12\,\sin^2\theta_W}}}$
& $\displaystyle{\frac{{\cal U}^*_{13}\,{\cal U}_{12}\, \cos^2\theta_W }{\sqrt{3-12\,\sin^2\theta_W}}}$  \\
&    &    \\ \hline
\hline
&     &        \\
$Z^{\prime} \bar d s$ &  
$\displaystyle{-\frac{{\cal V}^*_{12}\,{\cal V}_{11}\, \cos^2\theta_W }{\sqrt{3-12\,\sin^2\theta_W}}}$  &  $\displaystyle{\frac{{\cal V}^*_{12}\,{\cal V}_{11}\, \cos^2\theta_W }{\sqrt{3-12\,\sin^2\theta_W}}}$  \\
&      &      \\ \hline
\hline
&       &   \\
$Z^{\prime} \bar b d$  &   $\displaystyle{-\frac{{\cal V}^*_{13}\,{\cal V}_{11}\, \cos^2\theta_W }{\sqrt{3-12\,\sin^2\theta_W}}}$  &
$\displaystyle{\frac{{\cal V}^*_{13}\,{\cal V}_{11}\, \cos^2\theta_W }{\sqrt{3-12\,\sin^2\theta_W}}}$  \\
&     &     \\  \hline
\hline
&       &   \\
$Z^{\prime} \bar b s$  &   $\displaystyle{-\frac{{\cal V}^*_{13}\,{\cal V}_{12}\, \cos^2\theta_W }{\sqrt{3-12\,\sin^2\theta_W}}}$  &
$\displaystyle{\frac{{\cal V}^*_{13}\,{\cal V}_{12}\, \cos^2\theta_W }{\sqrt{3-12\,\sin^2\theta_W}}}$  \\
&     &     \\  \hline
\end{tabular}
\end{center}
\end{footnotesize}
\caption{The flavor-changing vector and axial-vector couplings to quarks ($u$- and $d$-type ) induced by $Z^{\prime}$ in the 3-3-1  model.}
\end{table}

\section{Results and Conclusions}

The asymmetry  for top quark pair  production is defined as 
$$A_{FB} \,  = \, \frac {N_t \left( \cos\, \theta \ \ge \, 0 \right) \, - \, N_{\bar t} \left( \cos\, \theta \ \ge \, 0 \right)}
{N_t \left( \cos\, \theta \ \ge \, 0 \right) \, + \, N_{\bar t} \left( \cos\, \theta \ \ge \, 0 \right)},$$
\nl where $\theta$ is the top quark momentum  direction with respect to the beam axis in the $ t \bar t$ rest frame. 

We use the CompHep package \cite{HEP} with CTEQ6  parton distribution functions, by fixing the top quark mass as $175$ GeV, and we adopt 
for $M_{Z^\prime}$ the values from $300$ to  $1000$ GeV;  the corresponding $Z^\prime$ widths are shown in the Table II.
In order to calculate the total cross section and to present the invariant mass and transverse momentum distributions, we have to fix some model
 parameters and to introduce kinematical cuts ($p_T \, > \,  20$ GeV and $\vert \eta \vert \, < \,  2$ for top and antitop) corresponding to the Tevatron conditions, where the proton antiproton collide at $\sqrt s = 1.96$ TeV. 

As discussed before, the {\it up}-type quark mixing leads to flavor-changing $Z^\prime$ coupling in both {\it s} and {\it t} channels. The elements of the 
mixing matrix are not fixed by experiments, and the values ${\cal U}_{11} = 0.933$ and ${\cal U}_{13} = 0.766$ have been used in the evaluation of the bilepton pair production cross section from the 3-3-1 model
\cite{YEB} in order to  respect the  unitarity constraint.  We have verified that it is possible to vary $U_{11}$ in the range $0.698-0.841$, keeping the correct elementary cross section behavior of the cited process, and that the result is not sensitive for $U_{13}$ variation. We present our results for two additional values for $U_{11}$ ($0.841$ and $0.698$). 
\vskip 0.5cm
  
Figures 1 and 2 present the invariant mass and transverse momentum distributions, for  $M_{Z^\prime} = 800$ GeV and $1000$ GeV, respectively, which are 
similar to the  SM ones, whereas for $M_{Z^\prime} = 500$ GeV the invariant mass  distribution presents some enhancement for $M_{tt}$ larger than $450$ GeV.

We display, in Figure 3,  the total cross section as a function of $M_{Z^\prime}$  for three values of $U_{11}$. We observe that  for $U_{11}=0.933$ it is possible to reproduce the experimental total cross section  result within $1 \sigma$ deviation, for  $M_{Z^\prime}$ from $350$ to $420$  
GeV and  from $525$ to $650$ GeV. Moreover, the tiny dependence of the total cross section with the  matrix mixing parameters is quite clear.

Next, we study the angular distribution in order to obtain the forward-backward asymmetry A$_{FB}$
 for top quark pair production. Our result is  shown in Figure 4. We observe that for $M_{Z^\prime}= \{350, \, 420\}$  GeV  and $U_{11}=0.841$ and $0.933$   the asymmetry is close with the central values measured by D$0$,  whereas for $M_{Z^\prime}= \{525, \, 650\}$ GeV with $U_{11}=0.814$ and $0.933$  it is possible to reproduce the CDF $A_{FB}$ measurement. For the mixing parameter  $U_{11} = 0.698$, the asymmetry is small for all studied $Z^\prime$ mass.  

 The calculated  $A_{FB}$, for two small ranges of $M_{Z^\prime}$ combined with the total cross section  and invariant mass distribution 
makes the 3-3-1 model a good candidate to describe  the discrepancy between the SM prediction and Fermilab result probably related to  new physics. 

\begin{figure}
\includegraphics[height=.5\textheight]{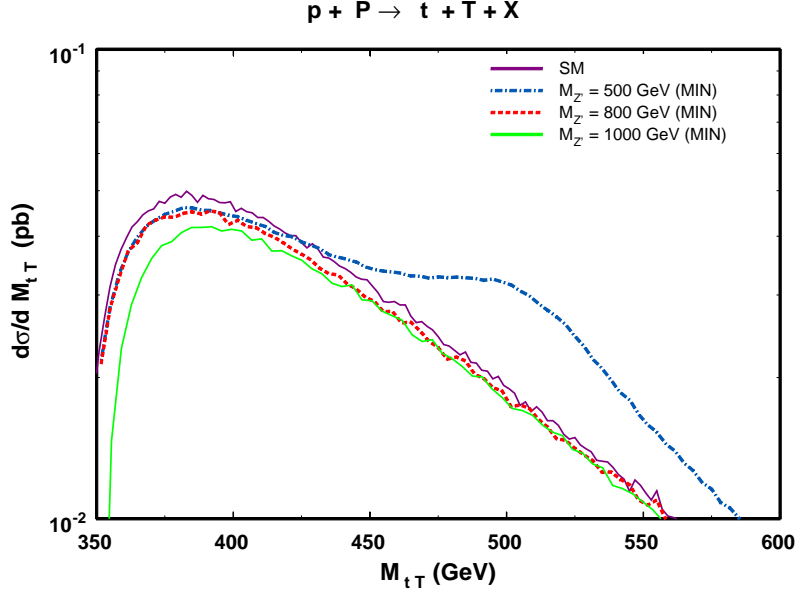}
\caption{\large Invariant mass distribution for the process $p + \bar p \longrightarrow t + \bar t + X$ ($\sqrt s = 1.96$ TeV) 
 for the 3-3-1 model considering some $M_{Z^{\prime}}$ values.}
\end{figure}
\begin{figure}
\includegraphics[height=.5\textheight]{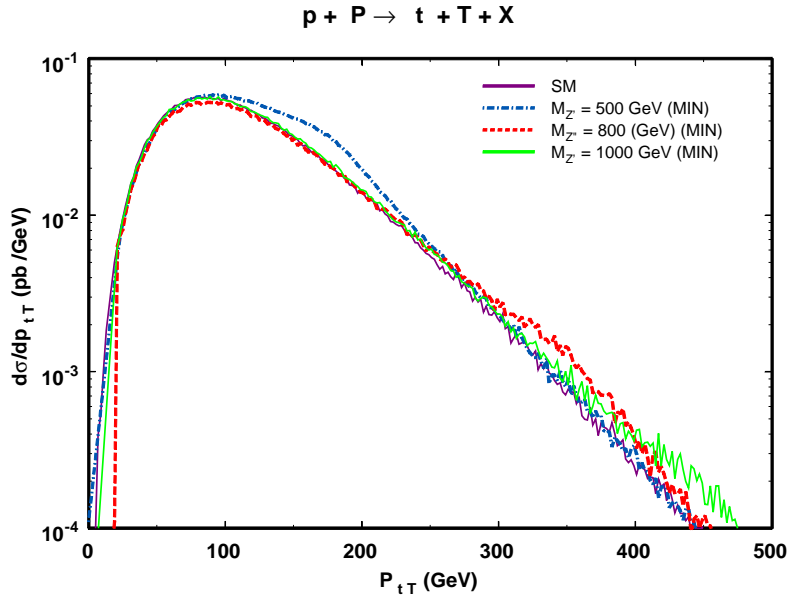}
\caption{\large Top transverse momentum distribution for the process $p + \bar p \longrightarrow 
t + \bar t + X$ ($\sqrt s = 1.96$ TeV) for the 3-3-1 model considering some $M_{Z^{\prime}}$ values.}
\end{figure}
\begin{figure}
\includegraphics[height=.5\textheight]{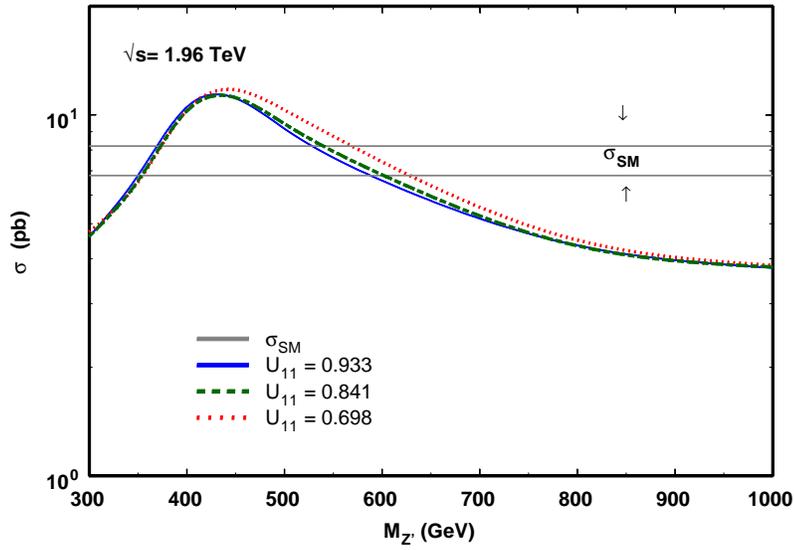}
\caption {The total cross section for the process $p + \bar p \longrightarrow 
t + \bar t + X$ ($\sqrt s = 1.96$ TeV) as a function of  $M_{Z^\prime}$ for three values of $U_{11}$ mixing parameters with $U_{13}$ fixed. The horizontal lines corresponds  to the SM value within $1 \sigma$ deviation.}
\end{figure}
\begin{figure}
\includegraphics[height=.5\textheight]{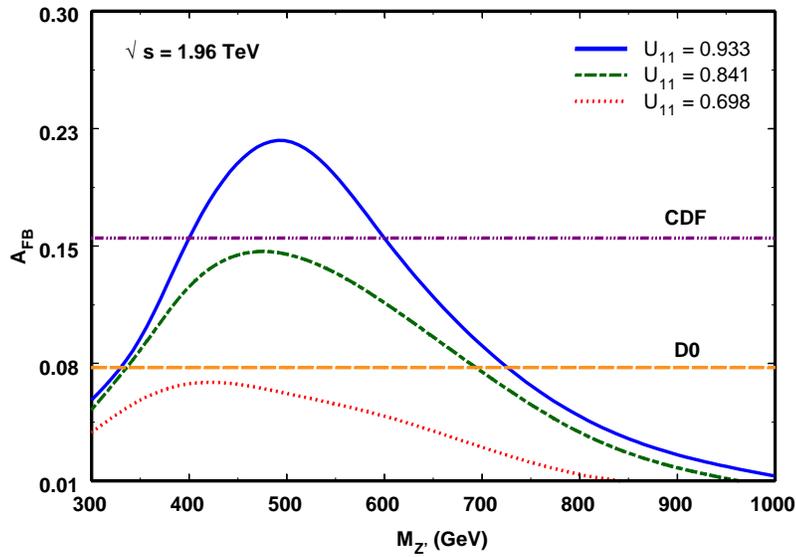}
\caption {The forward-backward asymmetry $A_{FB}$ for the process $p + \bar p \longrightarrow 
t + \bar t + X$ ($\sqrt s = 1.96$ TeV) as a function of  $M_{Z^\prime}$ for three values of $U_{11}$ mixing parameters with $U_{13}$ fixed. The horizontal lines correspond to the central values 
measured by the CDF and D$0$ Collaborations.}
\end{figure}

\begin{table}[h]\label{xuxa}
\begin{footnotesize}
\begin{center}
\begin{tabular}{||c|c|c|c|c|c|c|c|c||}
\hline
&    &    &    &    &  &    &  \\ 
$M_{Z^\prime}$ (GeV) & $300$ & $350$  & $ 400$  & $500$   & $600$ &  $800$  & $1000$ \\
&    &    &    &    &    &  &  \\ \hline
\hline
 &    &    &    &    &    &  & \\ 
$\Gamma_{Z^{\prime}}$ (GeV) & $30$ & $36$  & $43$  &  $57$  & $70$ & $97$ & $123$ \\
&    &    &    &    &   &   &  \\ \hline
\hline
\end{tabular}
\end{center}
\end{footnotesize}
\caption{The widths corresponding to different values for $M_{Z^\prime}$.}
\end{table}
\vskip 0.5 cm

\textit{Acknowledgments:} 
E. R. B. thanks Capes-PNPD. J. S. B. and Y. A. Coutinho thank FAPERJ for financial support.

\ed

\begin{table}[h]\label{xuxa}
\begin{footnotesize}
\begin{center}
\begin{tabular}{||c|c|c|c|c|c|c||}
\hline
&    &    &    &    &    &   \\ 
$M_{Z^\prime}$ (GeV) & $ 350 $ & $ 400 $ & $500$  & $ 600$   & $800$  & $1000$   \\
&    &    &   &  & & \\ \hline
\hline
&    &    &    &  & & \\
$A_{FB}$ ($\%$) & $0.13$  & $0.19$ & $0.23$ & $0.21$ &  $-0.01$  & $0.01$ 
 \\
&    &    &    &    &    & \\ \hline
\hline
&    &    &    &    &    & \\
$\sigma $ (pb) & $6.4$  & $9.2$ & $7.2$ & $4.7$ &  $2.8$  & $2.4$ 
 \\
&    &    &   &  &  & \\ \hline
\hline
\end{tabular}
\end{center}
\end{footnotesize}
\caption{The forward-backward asymmetry $A_{FB}$ and total cross section $\sigma$ corresponding to different values for $M_{Z^\prime}$}
\end{table}